\documentclass{article}
\usepackage[utf8]{inputenc}
\usepackage{arabtex}
\usepackage{textcomp}
\title{A novel terahertz time-domain spectroscopic endoscope based on a single photoconductive antenna chip}
\author{jitao Zhang\\ ECE Department,The University of Arizona, Tucson, AZ,85721\\
jitaozhang@email.arizona.edu }

\date{%
    \today
    \\[2\baselineskip]
    \normalfont\normalsize%
    \parbox{0.8\linewidth}{%
{\bfseries Abstract}: The common terahertz time-domain spectroscopy (THz-TDS) based on photoconductive antenna (PCA) needs two separate PCA chips. One PCA works as an emitter, and the other works as a receiver. For a reflection-type measurement, the technique called \emph{attenuated total reflection} usually is needed to enhance the reflection sensitivity. These make the system bulk and complicated for the reflection-type measurement. In this paper, we propose a novel THz-TDS endoscope that is specifically designed for reflection-type measurement. This THz-TDS endoscope is benefited from an \emph{integrated} photoconductive antenna (we call it\emph{i}PCA), which integrates the emitter and receiver on a single antenna chip. Therefore, the dimension of the endoscope can be shrunk as much as possible for practical usage. We present the design and working principle of this THz-TDS endoscope in details. It may open a promising way for the THz-TDS application in biomedical fields.
    }
}
\usepackage{natbib}
\usepackage{graphicx}

\begin{document}

\maketitle

\section{Introduction}

Terahertz time-domain spectroscopy (THz-TDS) is a technique that can characterize the sample's properties in THz band (usually from 0.1 to 3 THz)by coherently measuring its spectral response to the generated THz pulse in time-domain with the aid of ultrafast laser pulse. Owing to the various merits, such as measurement in a coherent way (can collect both amplitude and phase information simultaneously), high sensitivity and high temporal resolution,THz-TDS has been widely used in applications such as the spectroscopy and imaging. The photoconductive antenna (PCA) is the most commonly used device of THz pulse emitter and/or receiver in THz-TDS\citep{zhang2003}. Figure 1 demonstrates two typical manners (transmission-type and reflection-type) in which a THz-TDS works. 

\begin{figure}[h!]
\centering
\includegraphics[width=1\textwidth]{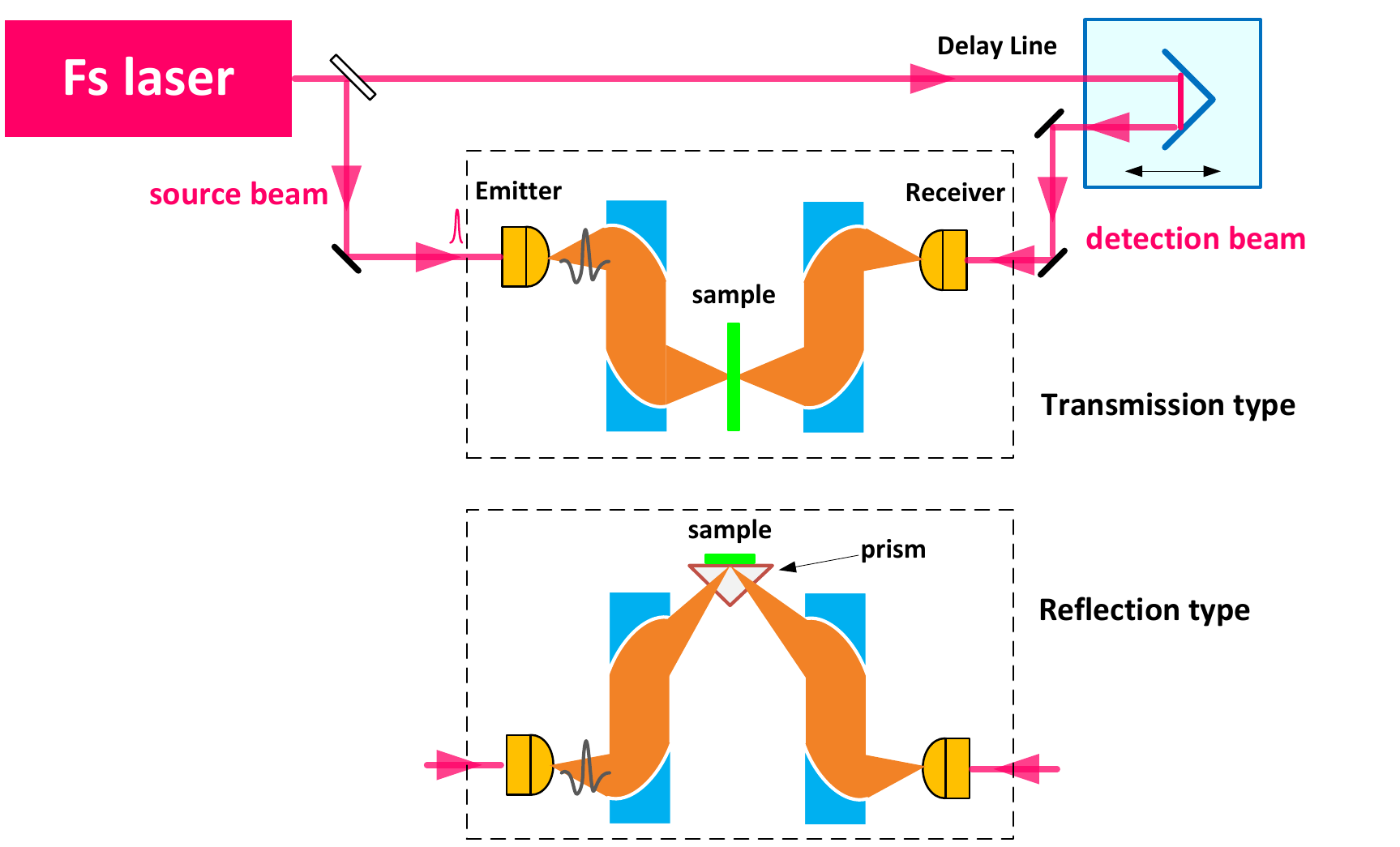}
\caption{Set-up of a common THz-TDS system for a transmission- and a reflection-type measurement}
\label{fig1:thz-tdssystem}
\end{figure}
The laser beam emitted from a femtosecond laser is splitted into a source beam and a detection beam. Two similar PCA chips are used as THz emitter and receiver. The THz pulse is generated on the back side of the PCA emitter when source beam is excited the PCA emitter. After transmitted or reflected by the sample, the temporal shape of the THz pulse is detected by the detection-beam-excited PCA receiver. The time delay between these two beams is scanned by a delay-line, which is usually an optical reflector driven by a motorized translation stage. The THz beam can be further reshaped by parabolic mirrors before measurement. For a sample that is transparent in THz region, a Transmission measurement can be employed. For a transmission measurement, the sample is usually placed at the focal plane of the focused THz beam, and the transmitted THz pulse is measured by the PCA receiver.However, when it comes to analyzing very lossy samples (e.g., aqueous biological and medical samples), the main drawback if most of the signal is lost due to high absorption. In this case a reflection measurement is preferred. In addition, to enhance the reflected signal,a technique called \emph{attenuated total reflection}(ATR)
is used by means of total internal reflection\citep{optlett2013}. As shown in Fig.1, the transparent high-resistivity silicon isosceles prism can be used to generate a total internal reflection at the interface between sample's lower surface and prism.The sample's surface and that underneath can be analyzed by THz-TDS in this way. However, the separated components of the system make it bulk and less flexible for biomedical application. There is a design that put two PCA antennas on the same silicon lens to fulfill the reflection measurement\citep{shen2011}. However, this will result in a non-common path of the emitter and receiver, so that the reflected THz beam cannot be collected and detected effectively.

In this work, we proposed a novel reflected-mode THz-TDS (THz-TDS endoscope). Owing to the novel antenna design, one single PCA chip can fulfill the needs for THz emission and detection, which is called as \emph{integrated} PCA (\emph{i}PCA). Since the emitter and the receiver share the same semiconductor material/substrate and silicon lens, the optical common path is satisfied inherently. This would enhance the collection efficiency of the reflected THz radiation. The use of this \emph{i}PCA makes the system very compact and flexible for reflection measurement. In addition, the system can work as a THz endoscope with the help of fiber-coupling design. This may open a new way for the THz application in biomedical field. The design of the antenna design and the THz-TDS endoscope is explained in Sec.2, and the working principle is demonstrated in Sec.3. A simple conclusion is finally made in Sec.4.

\section{THz-TDS endoscope design}
\subsection{Antenna design-\emph{i}PCA}
\begin{figure}[h!]
\centering
\includegraphics[width=0.8\textwidth]{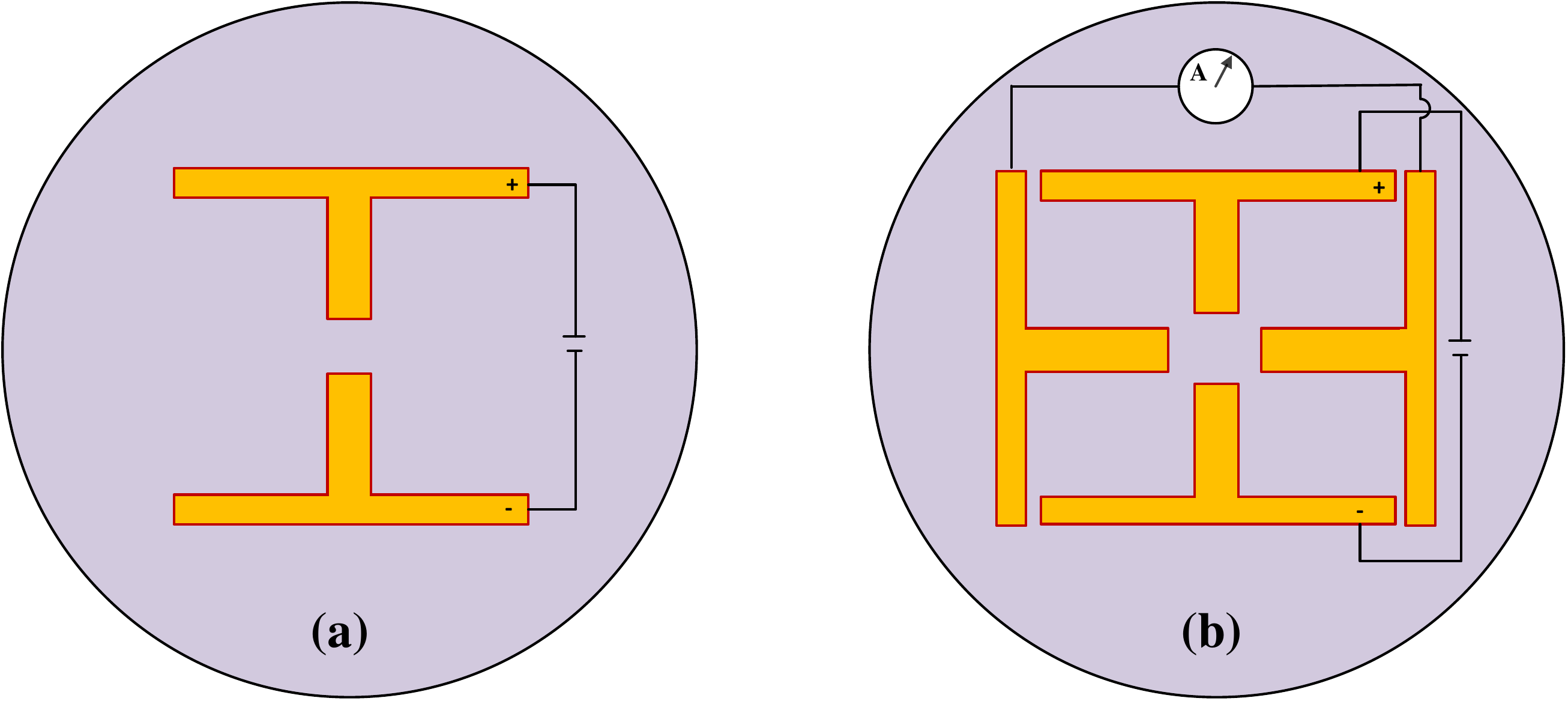}
\caption{PCA design for THz-TDS.(a)conventional PCA antenna.(b)\emph{i}PCA antenna integrated both the emitter and receiver.}
\label{fig2:pcadesign}
\end{figure}

The shape of a conventional PCA chip is shown in Fig.\ref{fig2:pcadesign}(a)\citep{austin1994}. 

A pair of metal electrodes working as THz-radiation antenna are deposited on the top surface of the active semiconductor (for example, low-temperature GaAs). The gap between the electrodes is shined by the laser pulse to generated photo-excited carriers (i.e. electron and hole) inside the semiconductor. If the electrodes are biased by a DC voltage, it will radiate THz wave owing to the transient flow of the photocurrent driven by the biased field. This is the way how a PCA emitter works. The same PCA chip can be used as THz receiver,where the electrodes are not biased but connected to a current meter. In the detection case, the incident THz radiation will drive the photo-excited carriers to generate transient current, which is proportional to the THz field. Therefore, the THz field can be recovered by the measured current. Usually, two separate PCA chips are used in a THz-TDS as emitter and receiver because they work in a different way, as shown in Fig. \ref{fig1:thz-tdssystem}. However, both the PCA emitter and receiver work on the same basis of the transient process of photo-excited carriers inside the semiconductor. This feature provides the possibility that realize both THz emitter and receiver on a single antenna chip. Suppose we deposit two pairs of electrodes on the same semiconductor (as shown in Fig.\ref{fig2:pcadesign}b), one pair is biased by a DC voltage, and the other is connected to a current meter. If we can make them work alternately, this single antenna chip can be an alternating THz emitter and receiver.

\subsection{THz-TDS endoscope design}
\begin{figure}[h!]
\centering
\includegraphics[width=0.8\textwidth]{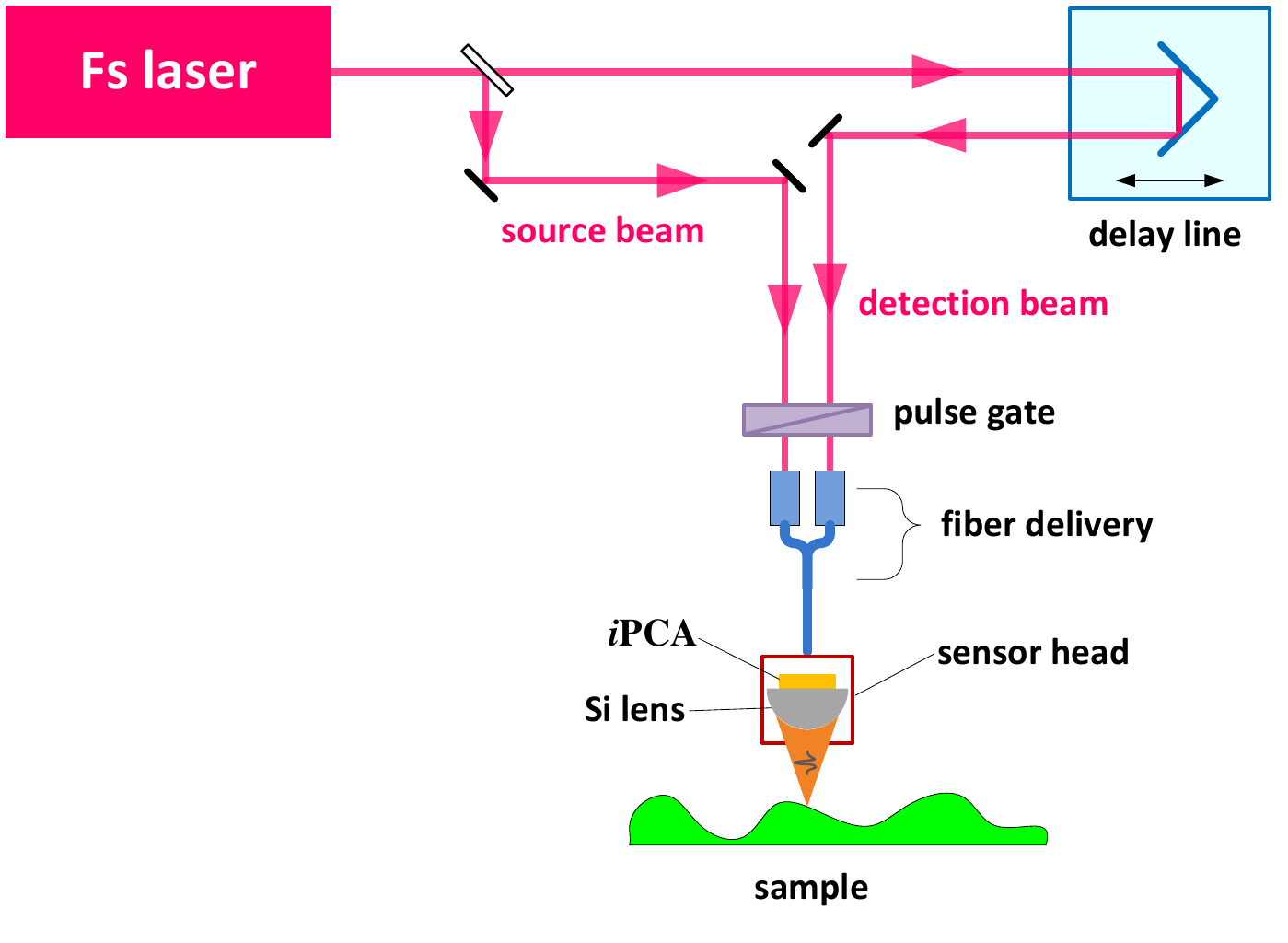}
\caption{Schematic design the THz-TDS endoscope}
\label{fig3:thz-tds-endoscope}
\end{figure}

The schematic design of the THz-TDS endoscope is shown in Fig.\ref{fig3:thz-tds-endoscope}.

The source and detection beams emitted from the femtosecond laser are coupled into fiber for delivery after they pass through a pulse gate in free space. The function of the pulse gate is to switch on/off the two beams with same frequency but $180\textsuperscript{\textdegree}$ phase difference. In other words, only one beam is allowed to pass within a time period. A 2$\times$1 fiber combiner is used to combine two beams together and deliver to the shining area (i.e. the gap between the electrodes)of the \emph{i}PCA inside the sensor head. The antenna chip is mounted on the back surface of a Si lens, which is used for THz beam shaping,thereby improves the efficiency of the radiation of the THz source as well as the detection of the reflected THz beam. The information of the sample surface can be collected by comparing the reflected THz beam with the reference.

\section{How the THz-TDS endoscope works}
\begin{figure}[h!]
\centering
\includegraphics[width=0.8\textwidth]{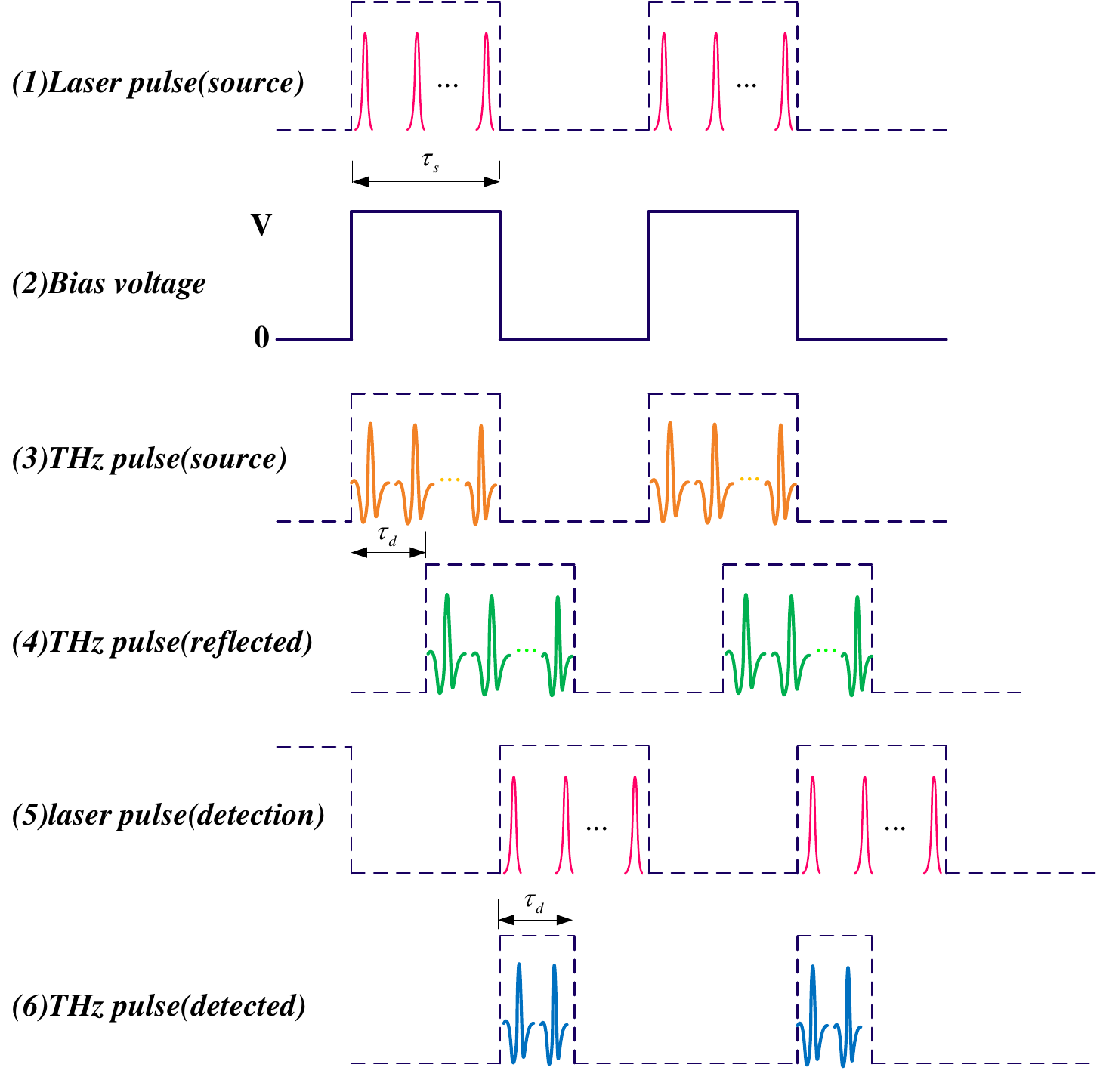}
\caption{Working procedure of the THz-TDS endoscope}
\label{fig4:procedure}
\end{figure}

The working procedure of this THz-TDS endoscope is shown in Fig.\ref{fig4:procedure}. As mentioned in Sec.2.2, the source and detection beams are modulated by the pulse gate with square wave. The bias voltage is modulated with the same manner by an electric switch. Both of them have a period of 2$\tau_{s}$. The same clock source should be used for the modulation of laser beam and bias voltage to make sure they are synchronous. 

When the source laser beam and bias voltage is switched on, the \emph{i}PCA works as a THz emitter. Because the repetition rate of the laser pulse is much higher than the modulation frequency, multiple THz pulses can be generated within the time window, and a discrete THz pulse-train can be emitted from the sensor head. After encountering the sample, This THz pulse-train will be reflected and recollected by the sensor head again after some time delay $\tau_{d}$(depends on the total path length it goes through). As the reflected THz beam arrives at the \emph{i}PCA chip, the detection laser beam is switched on but the source laser beam and bias voltage are switched off,thereby the \emph{i}PCA works as a THz receiver, and the reflected THz field will induce a flow of current, which is monitored by the current meter. The whole shape of the THz pulse is collected by scanning the delay line, as usually done in the typical THz-TDS. As long as $\tau_{d}\neq 2\emph{n}\tau_{s}$ (\emph{n} is a non-negative integer ), the reflected THz could be identified and measured. Even this is a case, it can be avoided by other ways. For example, we can either adjust the distance between the sensor head and the sample or tune the modulation frequency slightly.

\section{Conclusion}
We proposed a novel THz-TDS endoscope based on an \emph{i}PCA, which integrated both of the PCA emitter and receiver on a single antenna chip. The endoscope can implement reflection-type THz-TDS measurement with a very compact sensor head. This system can extend the application of THz-TDS in the biomedical fields. For example, it can be used for diagnosing the skin cancer and other epithelial tumours, and monitoring the decay of human tooth, et al.

\bibliographystyle{plain}
\bibliography{references}
\end{document}